\documentstyle[12pt]{article}

\catcode`\@=11
\@addtoreset{equation}{section}
\catcode`\@=12
\title{Diagram technique for perturbation theory calculations of the effective conductivity of two-dimensional systems}
\author{I.M. Khalatnikov$^{1,2,3}$ and A.Yu. Kamenshchik$^{1,2}$}
\date{}
\begin{document}
\maketitle
\hspace{-5mm}$^{1}$ 
L.D. Landau Institute for Theoretical Physics,
Russian Academy of Sciences, Kosygina str. 2, 117334
Moscow, Russia\\
$^{2}$ Landau Network - Centro Volta, Villa Olmo,
Via Cantoni 1, 22100 Como, Italy\\
$^{3}$ Tel Aviv University, Raymond and Sackler Faculty of 
Exact Sciences,
School of Physics and Astronomy,
Ramat Aviv, 69978, Israel

\begin{abstract}
The perturbation theory for calculation of the effective conductivity 
of the plane consisting of pieces of different conductivities 
is constructed and the convenient diagram technique for this 
perturbation theory is elaborated.  
It is shown that for the chessboard perturbative calculations 
give results
which are in agreement with the well-known formula $\sigma_{eff} =
\sqrt{\sigma_1\sigma_2}$. 
The components of the tensor of effective conductivity for the 
anisotropic three-color chessboard are calculated. It is shown 
that the isotropic (symmetric) part of effective conductivity calculated up to 
the sixth order of perturbation theory satisfies the Bruggeman effective medium equation for symmetric three-color structures with equally partitioned 
components. 
We also consider isotropic three-color chessboard with nonequal weights of colors. In this case the perturbation theory already in fourth order contradicts the results following from the Bruggeman equation for nonequal weights.
\end{abstract}  

 \\
\\
\section{INTRODUCTION}
The problem of calculation of the effective conductivity of 
composite matherials is of interest from theoretical and 
phenomenological 
points of view and has been attracting  the attention of theorists 
beginning 
in the nineteenth' century \cite{xix}. 
The most developed is the theory 
of the effective conductivity of the plane, 
because here one can
use the property of duality between conductivity and resistance, 
which is
typical only for the two-dimensional Ohm's law. 
This duality consists in the fact that the rotation of the plane on
$\pi/2$ simultaneously with interchange between the vectors of current
density $\vec{j}$ and the electric field $\vec{E}$ does not change 
Ohm's law.
The most interesting 
two-dimensional result is that by Dykhne \cite{Dykhne}, 
who has considered
the plane covered by regions with two 
different conductivities $\sigma_1$
and $\sigma_2$. If the distribution is stochastical and 
the statistical 
weights
of two conductivities are equal then one can show that 
\begin{equation}
\sigma_{eff} = \sqrt{\sigma_1\sigma_2}.
\label{Dykhne}
\end{equation} 
The same result will hold also for some regular two-conductivities
(for the sake of convenience we shall call them below two-color)
structures on the plane, in particular, for the chessboard
as was shown by Keller in his earlier paper \cite{Keller}. 

The main goal of this paper is the construction of the perturbation 
theory for calculation of the effective conductivity on the plane 
for an arbitrary distribution of the conductivity and the 
representation of the obtained formulae in a convenient graphical
form. As applications of this theory we rederive the formula of 
Keller--Dykhne for the case of two-color chessboard. Then, using our diagram 
technique  we calculate the components of the tensor of effective conductivity
for the three-color chessboard which represents an anisotropic structure
with equal weights.
It is shown that the isotropic part of the effective conductivity calculated up 
to the sixth order of perturbation theory satisfies the Bruggeman 
effective medium equation for symmetric three-color structures with equally partitioned components \cite{Brug}.

We consider also the isotropic three-color chessboard with nonequally partitioned components. It is shown that in this case the corresponding Bruggeman equation failed.  
 
Notice, that recently, numerical simulation of the conductivity of regular, isotropic equal-weighted three-color structures on the plane was performed
and analyzed in the paper \cite{Fel}. 
Three-component dielectric media were considered also in \cite{Emets}.

The structure of the paper is the following:
in the second section we shall construct the perturbation theory for the effective conductivity and present the convenient diagrammatic representation for this technique; the third section is devoted to the application of 
this technique to the deduction of the formula for the effective conductivity 
of the two-color chessboard; finally, in Sec. IV we apply perturbative technique for the calculation of components of the tensor of effective conductivity for the three-color two-dimensional chessboard and show that 
its isotropic part satisfies the Bruggeman equation, then we consider isotropic three-color chessboard with nonequal weights and show that fourth-order calculations for this case contradict  the predictions following from the corresponding Bruggeman equation.

\section{PERTURBATION THEORY FOR THE EFFECTIVE CONDUCTIVITY}
Now, let us begin constucting the perturbation theory for 
the effective conductivity. The idea of using the perturbative methods 
for the calculation of effective conductivity was considered before by 
C. Herring \cite{Herring} and also 
Dykhne \cite{Dykhne1} and Bergman \cite{Bergman}.
Notice that Herring has obtained the formula for the tensor of effective conductivity of a locally anisotropic medium in the second order \cite{Herring}, while considering locally isotropic mediums we develop the technique applicable to arbitrary orders of perturbation theory. 

Ohm's law for a locally isotropic medium has the following form:
\begin{equation}
\vec{j} = \sigma \vec{\nabla}\varphi,
\label{Ohm}
\end{equation}
where $\vec{j}$ is the current density, $\sigma$ is the 
conductivity, which generally depends on coordinates $(x,y)$ and 
$\varphi$ is an electric potential.
The charge conservation rule for a stationary distribution of 
currents reads
\begin{equation}
{\rm div} \vec{j} = \vec{\nabla}\cdot \vec{j} = 0.
\label{conserv}
\end{equation}   
Substituting Eq. (\ref{Ohm}) into Eq. (\ref{conserv}) one gets
\begin{equation}
\Delta \varphi + \vec{\nabla}\ln \sigma \cdot \vec{\nabla}\varphi = 0.
\label{Ohm1}
\end{equation}
Let us suppose now that
\begin{eqnarray}
&&\sigma(x,y) = 1 + \alpha(x,y), \nonumber \\
&&\langle \alpha \rangle = 0.
\label{perturb}
\end{eqnarray}
Here, the average value of $\sigma(x,y)$ is choosen equal to 1 
for the sake of convenience and $\alpha(x,y)$ is a small function 
whose average on the plane, denoted by 
$\langle \rangle$ is equal to zero.

Let us represent the potential $\varphi$ in the form 
\begin{equation}
\varphi = \vec{E} \cdot \vec{r} + \psi(x,y).
\label{perturb1}
\end{equation}
Here, the vector field $\vec{E}$ should be choosen in such a way 
to provide the fulfillment of the relation
\begin{equation}
\langle \vec{\nabla} \psi  \rangle = 0.
\label{condition}
\end{equation}
Now for isotropic distributions of conductivity one can introduce 
the effective conductivity by means of relation
\begin{equation}
\vec{J} = \sigma_{eff} \vec{E},
\label{effective}
\end{equation}
where $\vec{J} = \langle \vec{j}\rangle$.
For anisotropic distributions of $\alpha(x,y)$ the relation 
(\ref{effective}) becomes a tensor one:
\begin{equation}
J_i = \sigma_{eff ij} E_j,\ \  i,j = 1,2.
\label{effective1}
\end{equation}

Meanwhile, substituting the expression for $\varphi$ from 
(\ref{perturb1}) into Eq. (\ref{Ohm1}) one gets Laplace equation
\begin{equation}
\Delta \psi + \vec{\nabla}\ln(1+\alpha) \cdot (\vec{E}
+\vec{\nabla}\psi) = 0.
\label{Ohm2}
\end{equation}

Now it is convenient to introduce Fourier expansions for the 
small functions $\alpha$ and $\psi$:
\begin{equation}
\alpha = \sum_{\vec{k}} \alpha_{\vec{k}} \exp(i\vec{k}\cdot \vec{r}),
\label{Fourier}
\end{equation}
\begin{equation}
\psi = \sum_{\vec{k}} \psi_{\vec{k}} \exp(i\vec{k}\cdot \vec{r}).
\label{Fourier1}
\end{equation}
Substituting expansions (\ref{Fourier})--(\ref{Fourier1}) into 
Eq. (\ref{Ohm2}) one has
\begin{equation}
\vec{k}^2 \psi_{\vec{k}} = \left[\vec{\nabla}
\left(\alpha - \frac{\alpha^2}{2} + \frac{\alpha^3}{3} -
\frac{\alpha^4}{4} + \cdots\right)\cdot(\vec{E} + \vec{\nabla}\psi)\right]_{\vec{k}}.
\label{Ohm3}
\end{equation}
Resolving Eq. (\ref{Ohm3}) by iterations one can find a 
perturbative expansion for the function $\psi_{\vec{k}}$.  
Then, substititing the formulas found for  $\psi_{\vec{k}}$ into Eqs. (\ref{effective}) or (\ref{effective1}) one can find the 
components of the tensor of effective conductivity. 

Let us forget for a while about the condition (\ref{condition}), 
providing the correct definition of the macroscopic vector 
$\vec{E}$. One can write down the following formulas for the 
perturbative corrections to 
potenial $\psi_{\vec{k}}$ and for the isotropic part of the tensor 
of effective conductivity 
\begin{equation}
\sigma_{eff} \equiv \frac{1}{2} (\sigma_{xx} + \sigma_{yy})
\label{isotropic}
\end{equation} 
 \begin{equation}
\psi_{\vec{k} 1} = \frac{i \alpha_{k} (\vec{k}\cdot\vec{E})}
{\vec{k}^2},
\label{psi1}
\end{equation}
\begin{equation}
\sigma_{eff 2} = -\sum_{\vec{k}} \frac{(\vec{k}\cdot\vec{E})^2}
{\vec{E}^2 \vec{k}^2}
\alpha_{\vec{k}} \alpha_{-\vec{k}},
\label{sigma2}
\end{equation}
\begin{equation}
\psi_{\vec{k} 2} = -i\sum_{\vec{p}} \frac{(\vec{p}\cdot\vec{E})(\vec{p}\cdot\vec{k})}
{\vec{k}^2 \vec{p}^2}  
\alpha_{\vec{p}} \alpha_{\vec{p}-\vec{k}},
\label{psi2}
\end{equation}
\begin{equation}
\sigma_{eff 3} = \sum_{\vec{p},\vec{k}} 
\frac{(\vec{k}\vec{E})(\vec{p}\vec{E})(\vec{p}\vec{k})}
{\vec{k}^2 \vec{p}^2 \vec{E}^2}
\alpha_{\vec{p}}\alpha_{\vec{k}-\vec{p}}\alpha_{-\vec{k}},
\label{sigma3}
\end{equation}
and so on.

Making induction one can show that 
\begin{eqnarray}
&&\sigma_{eff n} = (-1)^{n+1}\sum{\vec{k}_1,\cdots,\vec{k}_{n-1}}
\frac{(\vec{k}_1\vec{E})(\vec{k}_1\vec{k}_2)\cdots
(\vec{k}_{n-2}\vec{k}_{n-1})(\vec{k}_{n-1}\vec{E})}
{\vec{E}^2 \vec{k}_1^2\cdots\vec{k}_{n-1}^2}\nonumber \\
&&\times\alpha_{\vec{k}_1}\alpha_{\vec{k}_2-\vec{k}_1}
\cdots\alpha_{\vec{k}_{n-1}-\vec{k}_{n-2}}\alpha_{-k_{n-1}}.
\label{sigman}
\end{eqnarray}
Thus all the corrections to effective conductivity have a 
''chain form'' and can be represented graphically as 
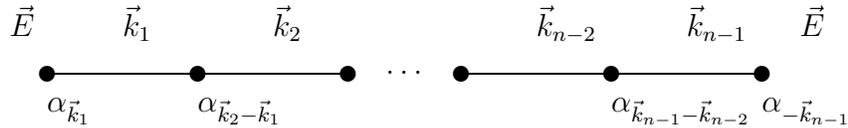
\setlength{\unitlength}{1mm}

\begin{picture}(120,20)
\put(5,15){$\vec{E}$}
\put(10,10){\circle*{2}}
\put(10,10){\line(4,0){20}}
\put(20,15){$\vec{k}_1$}
\put(30,10){\circle*{2}}
\put(30,5){$\alpha_{\vec{k}_2-\vec{k}_1}$}
\put(30,10){\line(4,0){20}}
\put(40,15){$\vec{k}_2$}
\put(50,10){\circle*{2}}
\put(55,10){$\ldots$}
\put(10,5){$\alpha_{\vec{k}_1}$}
\put(65,10){\circle*{2}}
\put(65,10){\line(4,0){20}}
\put(75,15){$\vec{k}_{n-2}$}
\put(85,10){\circle*{2}}
\put(85,5){$\alpha_{\vec{k}_{n-1}-\vec{k}_{n-2}}$}
\put(85,10){\line(4,0){20}}
\put(95,15){$\vec{k}_{n-1}$}
\put(105,10){\circle*{2}}
\put(105,5){$\alpha_{-\vec{k}_{n-1}}$}
\put(110,15){$\vec{E}$}
\end{picture}

However, to provide the condition (\ref{condition}) it is 
necessary to subtract from this diagram all the disconnected 
diagrams of $n$th order, consisting of 2 chains, then to add the 
diagrams, consisting of 3 chains and so on. The construction of such an alternating sum
corresponds to elimination of the terms containing the corrections, 
proportional to possible homogeneous 
vector harmonics $\vec{\nabla} \psi$.   
   
Now, we have to analyze in some detail the structure of expression (\ref{sigman}) and the corresponding diagram.
This diagram contains ''vertexes'', corresponding to
the coefficients of the Fourier expansion (\ref{Fourier}), 
and ''propagators'', corresponding to 
two-dimensional wave vectors $\vec{k}$.  Propagators
appear in 3 forms: 

\begin{picture}(120,40)
\put(10,30){\circle*{2}}
\put(10,30){\line(4,0){20}}
\put(30,30){\circle*{2}}
\put(35,29){$\equiv \frac12,$}
\put(55,30){\circle*{2}}
\put(55,30){\line(4,0){20}}
\put(65,32){$F$}
\put(75,30){\circle*{2}}
\put(80,29){$\equiv \frac12
\frac{k_x^2 - k_x^2}
{k_x^2 + k_x^2},$}
\put(10,10){\circle*{2}}
\put(10,10){\line(4,0){20}}
\put(20,12){$G$}
\put(30,10){\circle*{2}}
\put(35,9)
{$ \equiv \frac{k_x k_y}{k_x^2 + k_x^2}.$}
\end{picture}

It is remarkable that the formula (\ref{sigman}) 
(suplemented, as was explained earlier, 
by alternatingly subtracted terms)
contains information about all the elements of the tensor of
effective conductivity. Diagrams, corresponding to the 
isotropic part of this tensor given by the definition
(\ref{isotropic}) contain an even number of propagators 
$F$ and an even number of propagators $G$. Diagrams, 
corresponding to anisotropic part of the diagonal elements of this tensor
\begin{equation}
\sigma_{anysot} \equiv \frac12 (\sigma_{xx} - \sigma_{yy}),
\label{anysot}
\end{equation}
contain an odd number of propagators $F$ and an even number of $G$. 
Finally, diagrams, corresponding to non-diagonal elements of 
the effective conductivity tensor
$\sigma_{xy}$, 
contain an odd number of propagators $G$.

\section{PERTURBATIVE CALCULATIONS FOR THE TWO-COLOR CHESSBOARD}
Now, to show how this perturbation theory and diagram 
technique works, we shall consider the chessboard case.
Let us suppose that
\begin{eqnarray}
&&\sigma_1 = 1 + \delta,\nonumber \\
&&\sigma_2 = 1 - \delta.
\label{chess}
\end{eqnarray}
Then
\begin{equation}
\sigma_{eff} = \sqrt{1 - \delta^2} = 
1 -  \frac{\delta^2}{2} - \frac{\delta^4}{8}
-\frac{\delta^6}{16} - \frac{5 \delta^8}{128}
- \frac{7 \delta^{10}}{256} + \cdots.
\label{chess1}
\end{equation}

To reproduce this formula by means of perturbation theory, 
we should calculate the Fourier coefficients for the function 
$\alpha(x,y)$. Let us $\alpha(x,y)$ is a periodical function 
with periods $2\pi$ in both coordinates which behaves as 
\begin{eqnarray}
&&\alpha(x,y) = - \delta,\  if \   0<x<\pi,\ 0<y<\pi ; \nonumber \\    
&&\alpha(x,y) = + \delta,\  if \   \pi<x<2\pi,\ 0<y<\pi ; \nonumber \\    
&&\alpha(x,y) = + \delta,\  if \   0<x<\pi,\ \pi<y<2\pi ; \nonumber \\    
&&\alpha(x,y) = - \delta,\  if \   \pi<x<2\pi,\ \pi<y<2\pi.
\label{alpha}
\end{eqnarray}

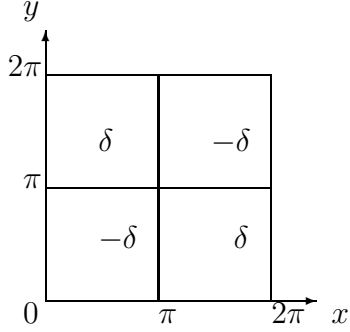
\begin{picture}(43,43)
\put(5,5){\vector(2,0){36}}
\put(5,5){\vector(0,2){36}}
\put(2,2){$0$}
\put(20,2){$\pi$}
\put(35,2){$2\pi$}
\put(5,20){\line(2,0){30}}
\put(5,35){\line(2,0){30}}
\put(20,5){\line(0,2){30}}
\put(35,5){\line(0,2){30}}
\put(2,20){$\pi$}
\put(0,35){$2\pi$}
\put(43,2){$x$}
\put(2,43){$y$}
\put(12,12){$-\delta$}
\put(30,12){$\delta$}
\put(12,25){$\delta$}
\put(27,25){$-\delta$}
\end{picture}

Then
\begin{equation}
\alpha_{2k,m} = \alpha_{n,2l} = 0,
\label{alpha1}
\end{equation}
\begin{equation}
\alpha_{2k+1,2l+1} = \frac{16\delta}
{(2\pi)^2(2k+1)(2l+1)}.
\label{alpha2}
\end{equation}

It is easy to see that all the diagrams with an odd number of 
vertices give zero contribution. 
Indeed, the sum of wave numbers of all the vertices of the 
diagram should be equal to zero (see Eq. (\ref{sigman})) and 
only odd values of wave numbers are possible 
(see Eqs. (\ref{alpha1})-(\ref{alpha2})). However, the sum of an
odd number of odd numbers cannot be equal to zero. 
Thus, we should consider only even orders of the perturbation theory.

Then, if the diagram under consideration contains an odd number 
of propagators $G$, the contribution of this 
diagram is equal to zero, because for any set of momenta
$\vec{k} = k_x,k_y$ there is the set $\vec{k}' = -k_x,k_y$, 
giving the contribution of the opposite sign.   
Analogously, the diagrams containing an odd number of 
propagators $F$ give vanishing contributions because, the 
contribution of any diagrams with a certain set of momenta 
$\vec{k} = k_x,k_y$ is cancelled by that with momenta 
$\vec{k} = k_y,k_x$. Thus, $\sigma_{anisot}$ and $\sigma_{xy}$ 
are equal to zero and the effective conductivity of the 
chessboard is isotropic.  

Now, we are in a position 
to calculate the diagrams corresponding 
to the effective conductivity of the chessboard. 
In the second order of 
the perturbation theory 
one has only the diagram 

\begin{picture}(100,10)
\put(10,5){\circle*{2}}
\put(10,5){\line(4,0){20}}
\put(30,5){\circle*{2}}
\end{picture}

which corresponds to the following expression:
\begin{eqnarray}
&&\sigma_{eff 2} = -\frac{1}{2} \sum_{\vec{k}} 
\alpha_{\vec{k}}\alpha_{-\vec{k}}\nonumber \\
&& = -\frac{256\delta^2}{2(2\pi)^4}\sum_{k=-\infty}^{\infty}
\sum_{l=-\infty}^{\infty}
\frac{1}{(2k+1)^2 (2l+1)^2}.
\label{second}
\end{eqnarray}
Using the well-known formula \cite{Abr}
\begin{equation}
\sum_{n=1}^{\infty} \frac{1}{n^2} = \zeta_R(2) = \frac{\pi^2}{6}
\label{Riemann}
\end{equation}
one has 
\begin{equation}
\sigma_{eff 2} = - \frac{\delta^2}{2}.
\label{second1}
\end{equation}

The set of diagrams of the fourth order has the following form:

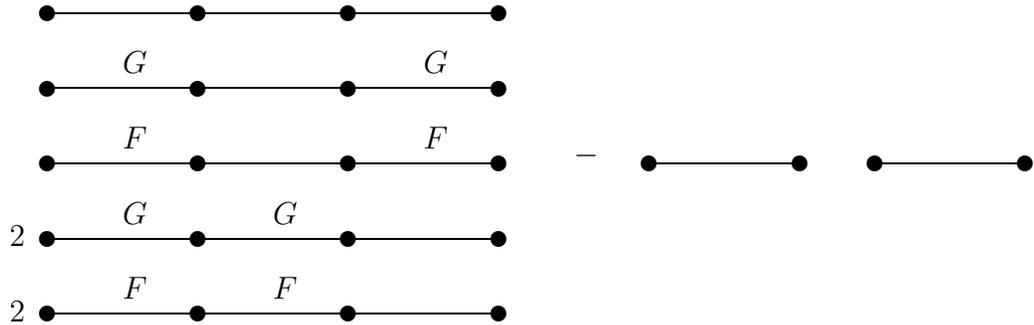
\begin{picture}(150,60)
\put(10,55){\circle*{2}}
\put(10,55){\line(4,0){20}}
\put(30,55){\circle*{2}}
\put(30,55){\line(4,0){20}}
\put(50,55){\circle*{2}}
\put(50,55){\line(4,0){20}}
\put(70,55){\circle*{2}}
\put(10,45){\circle*{2}}
\put(10,45){\line(4,0){20}}
\put(20,47){$G$}
\put(30,45){\circle*{2}}
\put(30,45){\line(4,0){20}}
\put(50,45){\circle*{2}}
\put(50,45){\line(4,0){20}}
\put(60,47){$G$}
\put(70,45){\circle*{2}}
\put(10,35){\circle*{2}}
\put(10,35){\line(4,0){20}}
\put(20,37){$F$}
\put(30,35){\circle*{2}}
\put(30,35){\line(4,0){20}}
\put(50,35){\circle*{2}}
\put(50,35){\line(4,0){20}}
\put(60,37){$F$}
\put(70,35){\circle*{2}}
\put(5,24){$2$}
\put(10,25){\circle*{2}}
\put(10,25){\line(4,0){20}}
\put(20,27){$G$}
\put(30,25){\circle*{2}}
\put(30,25){\line(4,0){20}}
\put(40,27){$G$}
\put(50,25){\circle*{2}}
\put(50,25){\line(4,0){20}}
\put(70,25){\circle*{2}}
\put(5,14){$2$}
\put(10,15){\circle*{2}}
\put(10,15){\line(4,0){20}}
\put(20,17){$F$}
\put(30,15){\circle*{2}}
\put(30,15){\line(4,0){20}}
\put(40,17){$F$}
\put(50,15){\circle*{2}}
\put(50,15){\line(4,0){20}}
\put(70,15){\circle*{2}}
\put(80,35){$-$}
\put(90,35){\circle*{2}}
\put(90,35){\line(4,0){20}}
\put(110,35){\circle*{2}}
\put(120,35){\circle*{2}}
\put(120,35){\line(4,0){20}}
\put(140,35){\circle*{2}}
\end{picture}

Here  the first connected diagram gives the contribution 
$\frac{\delta^4}{8}$
Combination of the second and third diagrams 
gives again  $\frac{\delta^4}{8}$, 
the disconnected diagram  gives the contribution 
$\frac{\delta^4}{8}$, while the last two connected diagrams 
do not give a contribution. Thus, the general result is equal to 
\begin{equation}
\sigma_{eff 4} = -\frac{\delta^4}{8}. 
\label{fourth1}
\end{equation}

In the following picture we shall present only the diagrams, 
giving nonvanishing contributions in the sixth order of 
perturbation theory with the corresponding results:

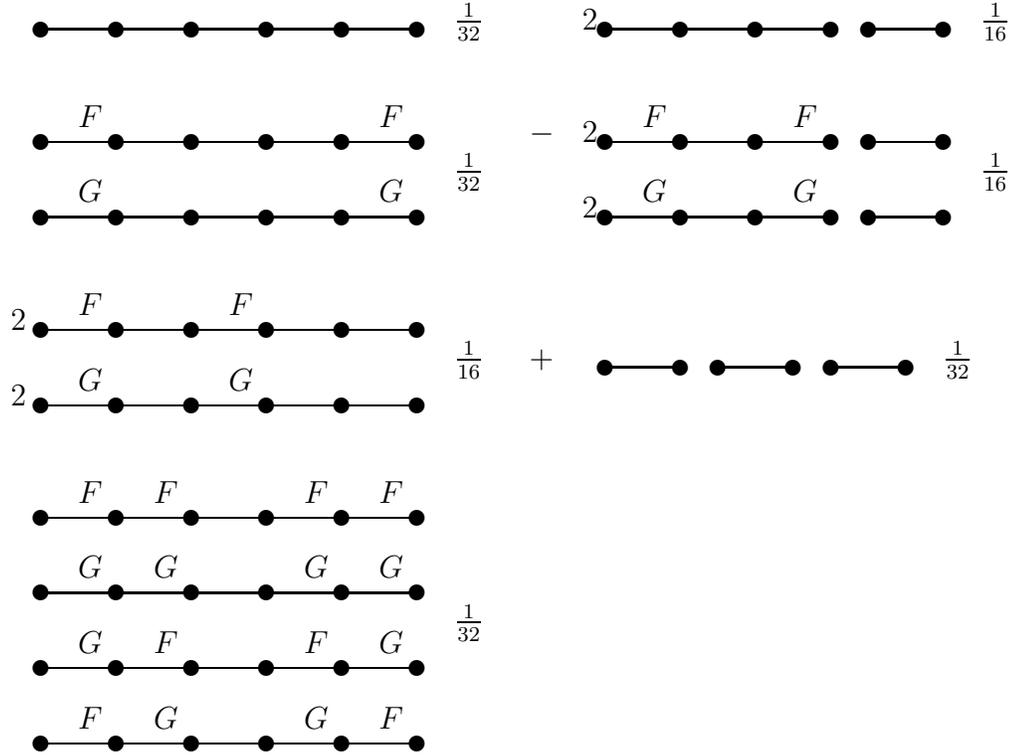
\begin{picture}(150,100)
\put(5,95){\circle*{2}}
\put(5,95){\line(4,0){10}}
\put(15,95){\circle*{2}}
\put(15,95){\line(4,0){10}}
\put(25,95){\circle*{2}}
\put(25,95){\line(4,0){10}}
\put(35,95){\circle*{2}}
\put(35,95){\line(4,0){10}}
\put(45,95){\circle*{2}}
\put(45,95){\line(4,0){10}}
\put(55,95){\circle*{2}}
\put(60,95){$\frac{1}{32}$}
\put(5,80){\circle*{2}}
\put(5,80){\line(4,0){10}}
\put(10,82){$F$}
\put(15,80){\circle*{2}}
\put(15,80){\line(4,0){10}}
\put(25,80){\circle*{2}}
\put(25,80){\line(4,0){10}}
\put(35,80){\circle*{2}}
\put(35,80){\line(4,0){10}}
\put(45,80){\circle*{2}}
\put(45,80){\line(4,0){10}}
\put(50,82){$F$}
\put(55,80){\circle*{2}}
\put(60,75){$\frac{1}{32}$}
\put(5,70){\circle*{2}}
\put(5,70){\line(4,0){10}}
\put(10,72){$G$}
\put(15,70){\circle*{2}}
\put(15,70){\line(4,0){10}}
\put(25,70){\circle*{2}}
\put(25,70){\line(4,0){10}}
\put(35,70){\circle*{2}}
\put(35,70){\line(4,0){10}}
\put(45,70){\circle*{2}}
\put(45,70){\line(4,0){10}}
\put(50,72){$G$}
\put(55,70){\circle*{2}}
\put(1,55){$2$}
\put(5,55){\circle*{2}}
\put(5,55){\line(4,0){10}}
\put(10,57){$F$}
\put(15,55){\circle*{2}}
\put(15,55){\line(4,0){10}}
\put(25,55){\circle*{2}}
\put(25,55){\line(4,0){10}}
\put(30,57){$F$}
\put(35,55){\circle*{2}}
\put(35,55){\line(4,0){10}}
\put(45,55){\circle*{2}}
\put(45,55){\line(4,0){10}}
\put(55,55){\circle*{2}}
\put(60,50){$\frac{1}{16}$}
\put(1,45){$2$}
\put(5,45){\circle*{2}}
\put(5,45){\line(4,0){10}}
\put(10,47){$G$}
\put(15,45){\circle*{2}}
\put(15,45){\line(4,0){10}}
\put(25,45){\circle*{2}}
\put(25,45){\line(4,0){10}}
\put(30,47){$G$}
\put(35,45){\circle*{2}}
\put(35,45){\line(4,0){10}}
\put(45,45){\circle*{2}}
\put(45,45){\line(4,0){10}}
\put(55,45){\circle*{2}}
\put(5,30){\circle*{2}}
\put(5,30){\line(4,0){10}}
\put(10,32){$F$}
\put(15,30){\circle*{2}}
\put(15,30){\line(4,0){10}}
\put(20,32){$F$}
\put(25,30){\circle*{2}}
\put(25,30){\line(4,0){10}}
\put(35,30){\circle*{2}}
\put(35,30){\line(4,0){10}}
\put(40,32){$F$}
\put(45,30){\circle*{2}}
\put(45,30){\line(4,0){10}}
\put(50,32){$F$}
\put(55,30){\circle*{2}}
\put(5,20){\circle*{2}}
\put(5,20){\line(4,0){10}}
\put(10,22){$G$}
\put(15,20){\circle*{2}}
\put(15,20){\line(4,0){10}}
\put(20,22){$G$}
\put(25,20){\circle*{2}}
\put(25,20){\line(4,0){10}}
\put(35,20){\circle*{2}}
\put(35,20){\line(4,0){10}}
\put(40,22){$G$}
\put(45,20){\circle*{2}}
\put(45,20){\line(4,0){10}}
\put(50,22){$G$}
\put(55,20){\circle*{2}}
\put(60,15){$\frac{1}{32}$}
\put(5,10){\circle*{2}}
\put(5,10){\line(4,0){10}}
\put(10,12){$G$}
\put(15,10){\circle*{2}}
\put(15,10){\line(4,0){10}}
\put(20,12){$F$}
\put(25,10){\circle*{2}}
\put(25,10){\line(4,0){10}}
\put(35,10){\circle*{2}}
\put(35,10){\line(4,0){10}}
\put(40,12){$F$}
\put(45,10){\circle*{2}}
\put(45,10){\line(4,0){10}}
\put(50,12){$G$}
\put(55,10){\circle*{2}}
\put(5,0){\circle*{2}}
\put(5,0){\line(4,0){10}}
\put(10,2){$F$}
\put(15,0){\circle*{2}}
\put(15,0){\line(4,0){10}}
\put(20,2){$G$}
\put(25,0){\circle*{2}}
\put(25,0){\line(4,0){10}}
\put(35,0){\circle*{2}}
\put(35,0){\line(4,0){10}}
\put(40,2){$G$}
\put(45,0){\circle*{2}}
\put(45,0){\line(4,0){10}}
\put(50,2){$F$}
\put(55,0){\circle*{2}}
\put(70,80){${\bf -}$}
\put(77,95){$2$}
\put(80,95){\circle*{2}}
\put(80,95){\line(4,0){10}}
\put(90,95){\circle*{2}}
\put(90,95){\line(4,0){10}}
\put(100,95){\circle*{2}}
\put(100,95){\line(4,0){10}}
\put(110,95){\circle*{2}}
\put(115,95){\circle*{2}}
\put(115,95){\line(4,0){10}}
\put(125,95){\circle*{2}}
\put(130,95){$\frac{1}{16}$}
\put(77,80){$2$}
\put(80,80){\circle*{2}}
\put(80,80){\line(4,0){10}}
\put(85,82){$F$}
\put(90,80){\circle*{2}}
\put(90,80){\line(4,0){10}}
\put(100,80){\circle*{2}}
\put(100,80){\line(4,0){10}}
\put(105,82){$F$}
\put(110,80){\circle*{2}}
\put(115,80){\circle*{2}}
\put(115,80){\line(4,0){10}}
\put(125,80){\circle*{2}}
\put(77,70){$2$}
\put(80,70){\circle*{2}}
\put(80,70){\line(4,0){10}}
\put(85,72){$G$}
\put(90,70){\circle*{2}}
\put(90,70){\line(4,0){10}}
\put(100,70){\circle*{2}}
\put(100,70){\line(4,0){10}}
\put(105,72){$G$}
\put(110,70){\circle*{2}}
\put(115,70){\circle*{2}}
\put(115,70){\line(4,0){10}}
\put(125,70){\circle*{2}}
\put(130,75){$\frac{1}{16}$}
\put(70,50){${\bf +}$}
\put(80,50){\circle*{2}}
\put(80,50){\line(4,0){10}}
\put(90,50){\circle*{2}}
\put(95,50){\circle*{2}}
\put(95,50){\line(4,0){10}}
\put(105,50){\circle*{2}}
\put(110,50){\circle*{2}}
\put(110,50){\line(4,0){10}}
\put(120,50){\circle*{2}}
\put(125,50){${\frac{1}{32}}$}

\end{picture}
\vskip5mm

Making summation, one has:
\begin{equation}
\sigma_{eff 6} = -\frac{\delta^6}{16}. 
\label{sixth1}
\end{equation}
Quite analogously, one can get also
\begin{equation}
\sigma_{eff 8} = -\frac{5\delta^8}{128} 
\label{eighth1}
\end{equation}
and 
\begin{equation}
\sigma_{eff 10} = -\frac{7\delta^{10}}{256}. 
\label{tenth1}
\end{equation}
One can see that the results of 
perturbative calculations (\ref{second1})-(\ref{tenth1}) 
coincide with those obtained by expansion of the general formula 
(\ref{chess1}). It confirms the reliability of 
the perturbation theory developed here and encourage us to apply it to
more complicated situations.

\section{PERTURBATIVE CALCULATIONS FOR THE THREE-COLOR CHESSBOARDS}
Now we would like to consider three-color structures with square lattice symmetry, which we call the three-color chessboards.
To do it, let us turn again to the formula (\ref{Dykhne}).
It is easy to notice that Eq. (\ref{Dykhne}) could be 
rewritten in the following form:
\begin{equation}
\frac{\sigma_{eff} - \sigma_1}{\sigma_{eff} + \sigma_1} +
\frac{\sigma_{eff} - \sigma_2}{\sigma_{eff} + \sigma_2} = 0.
\label{two-col}
\end{equation}
The equation (\ref{two-col}) was written by Bruggeman \cite{Brug}
as an 
approximative one (in the dipole approximation), but appears to be exact for the large class of
two-color coverings \cite{Dykhne}.
One can try to generalize this equation for the case of 
three conductivities (three-color system) with equal 
weights
\begin{equation}
\frac{\sigma_{eff} - \sigma_1}{\sigma_{eff} + \sigma_1} +
\frac{\sigma_{eff} - \sigma_2}{\sigma_{eff} + \sigma_2} +
\frac{\sigma_{eff} - \sigma_3}{\sigma_{eff} + \sigma_3} = 0,
\label{three-col}
\end{equation}
which is equivalent to 
\begin{eqnarray} 
&&\sigma_{eff}^3 + \frac{1}{3}(\sigma_1+\sigma_2+\sigma_3)
\sigma_{eff}^2 \nonumber \\
&&- \frac{1}{3}(\sigma_1\sigma_2 + \sigma_1\sigma_2 
+ \sigma_2\sigma_3)\sigma_{eff} - \sigma_1\sigma_2\sigma_3
= 0.
\label{three-col1}
\end{eqnarray}

There are different tesselations of the plane allowing 
three-color (i.e. three-conductivities) covering
 (see, for example, Ref. \cite{tessel}).
Here we shall consider the simplest tesselation: it is again 
a chessboard but covered by three colors with the same 
statistical weights. We shall introduce again the function 
$\alpha(x,y)$ which is periodic with periods $2\pi$ and is 
defined by the following rules:
\begin{eqnarray}
&&\alpha(x,y) = \delta,\  if\  
0<x<\frac{2\pi}{3},0<y<\frac{2\pi}{3};\nonumber \\      
&&\alpha(x,y) = 0,\  if\  \frac{2\pi}{3}<x<\frac{4\pi}{3},
0<y<\frac{2\pi}{3};\nonumber \\      
&&\alpha(x,y) = -\delta,\  if\  \frac{4\pi}{3}<x<2\pi, 0<y<\frac{2\pi}{3};\nonumber \\
&&\alpha(x,y) = -\delta,\  if\  0<x<\frac{2\pi}{3},\frac{2\pi}{3}<y<\frac{4\pi}{3};
\nonumber \\
&&\alpha(x,y) = \delta,\  if\  \frac{2\pi}{3}<x<\frac{4\pi}{3},
\frac{2\pi}{3}<y<\frac{4\pi}{3};\nonumber \\
&&\alpha(x,y) = 0,\  
if\  \frac{4\pi}{3}<x<2\pi,\frac{2\pi}{3}<y<\frac{4\pi}{3};
\nonumber \\
&&\alpha(x,y) = 0,\  if\  0<x<\frac{2\pi}{3},\frac{4\pi}{3}<y<2\pi;
\nonumber \\
&&\alpha(x,y) = -\delta,\  if\  \frac{2\pi}{3}<x<\frac{4\pi}{3},\frac{4\pi}{3}<y<2\pi;
\nonumber \\
&&\alpha(x,y) = \delta,\  if\ 
\frac{4\pi}{3}<x<2\pi,\frac{4\pi}{3}<y<2\pi.
\label{three-color}
\end{eqnarray}

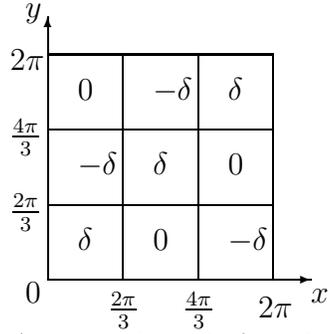
\begin{picture}(45,45)
\put(5,5){\vector(2,0){35}}
\put(5,5){\vector(0,2){35}}
\put(2,2){$0$}
\put(0,13){$\frac{2\pi}{3}$}
\put(0,23){$\frac{4\pi}{3}$}
\put(0,33){$2\pi$}
\put(2,40){$y$}
\put(5,15){\line(2,0){30}}
\put(5,25){\line(2,0){30}}
\put(5,35){\line(2,0){30}}
\put(15,5){\line(0,2){30}}
\put(25,5){\line(0,2){30}}
\put(35,5){\line(0,2){30}}
\put(13,0){$\frac{2\pi}{3}$}
\put(23,0){$\frac{4\pi}{3}$}
\put(33,0){$2\pi$}
\put(40,2){$x$}
\put(9,9){$\delta$}
\put(19,9){$0$}
\put(29,9){$-\delta$}
\put(9,19){$-\delta$}
\put(19,19){$\delta$}
\put(29,19){$0$}
\put(9,29){$0$}
\put(19,29){$-\delta$}
\put(29,29){$\delta$}
\end{picture}

Apparently, this function corresponds to the periodic distribution 
of three conductivitites

\begin{eqnarray}
&&\sigma_1 = 1 + \delta, \nonumber \\
&&\sigma_2 = 1 - \delta, \nonumber \\
&&\sigma_3 = 1.
\label{three-color1}
\end{eqnarray}
Correspondingly, the Fourier coefficients have the following form:
\begin{equation}
\alpha_{3k,m} = \alpha_{n,3l} = 0,
\label{three-color2}
\end{equation}
\begin{equation}
\alpha_{3k+1,3l+1} = \alpha_{3k+2,3l+2} = 0,
\label{three-color3}
\end{equation}
\begin{equation}
\alpha_{3k+2,3l+1} = \alpha^*_{3l+1,3k+2} 
= \frac{9\sqrt{3}ie^{i\pi/3}}{(2\pi)^2(3k+2)(3l+1)}.
\label{three-color4}
\end{equation}

Now, we are in a position to calculate perturbative contributions 
to the isotropic part of the effective conductivity for 
three-color chessboard. In the second order of perturbation theory 
we have again only one diagram, the contribution of which is equal to 
\begin{equation}
\sigma_{eff 2} = -\frac{\delta^2}{3}.
\label{second3}
\end{equation}
(Here again the formula (\ref{Riemann}) was used).
In the fourth order we have the following set of diagrams which have nonvanishing contributions:

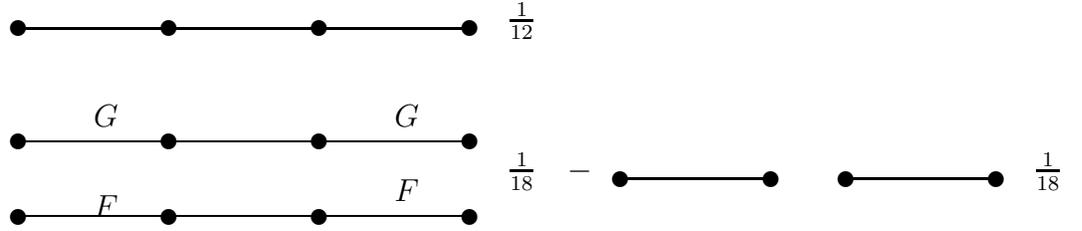
\begin{picture}(150,60)
\put(10,55){\circle*{2}}
\put(10,55){\line(4,0){20}}
\put(30,55){\circle*{2}}
\put(30,55){\line(4,0){20}}
\put(50,55){\circle*{2}}
\put(50,55){\line(4,0){20}}
\put(70,55){\circle*{2}}
\put(75,55){$\frac{1}{12}$}
\put(10,40){\circle*{2}}
\put(10,40){\line(4,0){20}}
\put(20,42){$G$}
\put(30,40){\circle*{2}}
\put(30,40){\line(4,0){20}}
\put(50,40){\circle*{2}}
\put(50,40){\line(4,0){20}}
\put(60,42){$G$}
\put(75,35){$\frac{1}{18}$}
\put(70,40){\circle*{2}}
\put(10,30){\circle*{2}}
\put(10,30){\line(4,0){20}}
\put(20,30){$F$}
\put(30,30){\circle*{2}}
\put(30,30){\line(4,0){20}}
\put(50,30){\circle*{2}}
\put(50,30){\line(4,0){20}}
\put(60,32){$F$}
\put(70,30){\circle*{2}}
\put(83,35){${\bf -}$}
\put(90,35){\circle*{2}}
\put(90,35){\line(4,0){20}}
\put(110,35){\circle*{2}}
\put(120,35){\circle*{2}}
\put(120,35){\line(4,0){20}}
\put(140,35){\circle*{2}}
\put(145,35){$\frac{1}{18}$}
\end{picture}
\vskip-27mm

which result in 
\begin{equation}
\sigma_{eff 4} = -\frac{\delta^4}{12}.
\label{fourth3}
\end{equation}
Calculating the sums corresponding to the diagrams presented 
above we have used the formula \cite{Abr}
\begin{equation}
S \equiv \sum_{k=-\infty}^{\infty}
\frac{1}{3k+1} = -\sum_{k=-\infty}^{\infty}
\frac{1}{3k+2} =
-\sum_{k=0}^{\infty}
\frac{1}{(3k+1)(3k+2)} = \frac{\pi}{3\sqrt{3}}.
\label{sum}
\end{equation}

Below we present the set of all the sixth-order diagrams 
giving nonvanishing contributions:

\begin{picture}(150,130)
\put(5,125){\circle*{2}}
\put(5,125){\line(4,0){10}}
\put(15,125){\circle*{2}}
\put(15,125){\line(4,0){10}}
\put(25,125){\circle*{2}}
\put(25,125){\line(4,0){10}}
\put(35,125){\circle*{2}}
\put(35,125){\line(4,0){10}}
\put(45,125){\circle*{2}}
\put(45,125){\line(4,0){10}}
\put(55,125){\circle*{2}}
\put(60,125){$\frac{1}{48}$}
\put(5,110){\circle*{2}}
\put(5,110){\line(4,0){10}}
\put(10,112){$F$}
\put(15,110){\circle*{2}}
\put(15,110){\line(4,0){10}}
\put(25,110){\circle*{2}}
\put(25,110){\line(4,0){10}}
\put(35,110){\circle*{2}}
\put(35,110){\line(4,0){10}}
\put(45,110){\circle*{2}}
\put(45,110){\line(4,0){10}}
\put(50,112){$F$}
\put(55,110){\circle*{2}}
\put(60,105){$\frac{1}{72}$}
\put(5,100){\circle*{2}}
\put(5,100){\line(4,0){10}}
\put(10,102){$G$}
\put(15,100){\circle*{2}}
\put(15,100){\line(4,0){10}}
\put(25,100){\circle*{2}}
\put(25,100){\line(4,0){10}}
\put(35,100){\circle*{2}}
\put(35,100){\line(4,0){10}}
\put(45,100){\circle*{2}}
\put(45,100){\line(4,0){10}}
\put(50,102){$G$}
\put(55,100){\circle*{2}}
\put(1,85){$2$}
\put(5,85){\circle*{2}}
\put(5,85){\line(4,0){10}}
\put(10,87){$F$}
\put(15,85){\circle*{2}}
\put(15,85){\line(4,0){10}}
\put(25,85){\circle*{2}}
\put(25,85){\line(4,0){10}}
\put(30,87){$F$}
\put(35,85){\circle*{2}}
\put(35,85){\line(4,0){10}}
\put(45,85){\circle*{2}}
\put(45,85){\line(4,0){10}}
\put(55,85){\circle*{2}}
\put(60,80){$\frac{1}{36}$}
\put(1,75){$2$}
\put(5,75){\circle*{2}}
\put(5,75){\line(4,0){10}}
\put(10,77){$G$}
\put(15,75){\circle*{2}}
\put(15,75){\line(4,0){10}}
\put(25,75){\circle*{2}}
\put(25,75){\line(4,0){10}}
\put(30,77){$G$}
\put(35,75){\circle*{2}}
\put(35,75){\line(4,0){10}}
\put(45,75){\circle*{2}}
\put(45,75){\line(4,0){10}}
\put(55,75){\circle*{2}}
\put(5,60){\circle*{2}}
\put(5,60){\line(4,0){10}}
\put(10,62){$F$}
\put(15,60){\circle*{2}}
\put(15,60){\line(4,0){10}}
\put(20,62){$F$}
\put(25,60){\circle*{2}}
\put(25,60){\line(4,0){10}}
\put(35,60){\circle*{2}}
\put(35,60){\line(4,0){10}}
\put(40,62){$F$}
\put(45,60){\circle*{2}}
\put(45,60){\line(4,0){10}}
\put(50,62){$F$}
\put(55,60){\circle*{2}}
\put(5,50){\circle*{2}}
\put(5,50){\line(4,0){10}}
\put(10,52){$G$}
\put(15,50){\circle*{2}}
\put(15,50){\line(4,0){10}}
\put(20,52){$G$}
\put(25,50){\circle*{2}}
\put(25,50){\line(4,0){10}}
\put(35,50){\circle*{2}}
\put(35,50){\line(4,0){10}}
\put(40,52){$G$}
\put(45,50){\circle*{2}}
\put(45,50){\line(4,0){10}}
\put(50,52){$G$}
\put(55,50){\circle*{2}}
\put(60,45){$\frac{1}{108}$}
\put(5,40){\circle*{2}}
\put(5,40){\line(4,0){10}}
\put(10,42){$G$}
\put(15,40){\circle*{2}}
\put(15,40){\line(4,0){10}}
\put(20,42){$F$}
\put(25,40){\circle*{2}}
\put(25,40){\line(4,0){10}}
\put(35,40){\circle*{2}}
\put(35,40){\line(4,0){10}}
\put(40,42){$F$}
\put(45,40){\circle*{2}}
\put(45,40){\line(4,0){10}}
\put(50,42){$G$}
\put(55,40){\circle*{2}}
\put(5,30){\circle*{2}}
\put(5,30){\line(4,0){10}}
\put(10,32){$F$}
\put(15,30){\circle*{2}}
\put(15,30){\line(4,0){10}}
\put(20,32){$G$}
\put(25,30){\circle*{2}}
\put(25,30){\line(4,0){10}}
\put(35,30){\circle*{2}}
\put(35,30){\line(4,0){10}}
\put(40,32){$G$}
\put(45,30){\circle*{2}}
\put(45,30){\line(4,0){10}}
\put(50,32){$F$}
\put(55,30){\circle*{2}}
\put(70,110){${\bf -}$}
\put(77,125){$2$}
\put(80,125){\circle*{2}}
\put(80,125){\line(4,0){10}}
\put(90,125){\circle*{2}}
\put(90,125){\line(4,0){10}}
\put(100,125){\circle*{2}}
\put(100,125){\line(4,0){10}}
\put(110,125){\circle*{2}}
\put(115,125){\circle*{2}}
\put(115,125){\line(4,0){10}}
\put(125,125){\circle*{2}}
\put(77,115){$2$}
\put(80,115){\circle*{2}}
\put(80,115){\line(4,0){10}}
\put(85,117){$F$}
\put(90,115){\circle*{2}}
\put(90,115){\line(4,0){10}}
\put(100,115){\circle*{2}}
\put(100,115){\line(4,0){10}}
\put(105,117){$F$}
\put(110,115){\circle*{2}}
\put(115,115){\circle*{2}}
\put(115,115){\line(4,0){10}}
\put(125,115){\circle*{2}}
\put(130,115){$\frac{5}{108}$}
\put(77,105){$2$}
\put(80,105){\circle*{2}}
\put(80,105){\line(4,0){10}}
\put(85,107){$G$}
\put(90,105){\circle*{2}}
\put(90,105){\line(4,0){10}}
\put(100,105){\circle*{2}}
\put(100,105){\line(4,0){10}}
\put(105,107){$G$}
\put(110,105){\circle*{2}}
\put(115,105){\circle*{2}}
\put(115,105){\line(4,0){10}}
\put(125,105){\circle*{2}}
\put(70,80){${\bf +}$}
\put(80,80){\circle*{2}}
\put(80,80){\line(4,0){10}}
\put(90,80){\circle*{2}}
\put(95,80){\circle*{2}}
\put(95,80){\line(4,0){10}}
\put(105,80){\circle*{2}}
\put(110,80){\circle*{2}}
\put(110,80){\line(4,0){10}}
\put(120,80){\circle*{2}}
\put(125,80){$\frac{1}{108}$}
\put(5,15){\circle*{2}}
\put(5,15){\line(4,0){10}}
\put(15,15){\circle*{2}}
\put(15,15){\line(4,0){10}}
\put(20,17){$F$}
\put(25,15){\circle*{2}}
\put(25,15){\line(4,0){10}}
\put(35,15){\circle*{2}}
\put(35,15){\line(4,0){10}}
\put(40,17){$F$}
\put(45,15){\circle*{2}}
\put(45,15){\line(4,0){10}}
\put(55,15){\circle*{2}}
\put(60,10){$\frac{1}{216}$}
\put(5,5){\circle*{2}}
\put(5,5){\line(4,0){10}}
\put(15,5){\circle*{2}}
\put(15,5){\line(4,0){10}}
\put(20,7){$G$}
\put(25,5){\circle*{2}}
\put(25,5){\line(4,0){10}}
\put(35,5){\circle*{2}}
\put(35,5){\line(4,0){10}}
\put(40,7){$G$}
\put(45,5){\circle*{2}}
\put(45,5){\line(4,0){10}}
\put(55,5){\circle*{2}}
\end{picture}

Making summation of these contributions, one has
\begin{equation}
\sigma_{eff 6} = -\frac{17\delta^6}{432}.
\label{sixth3}
\end{equation}

Thus, we have calculated a symmetric part of the tensor of effective conductivity $\sigma_{eff} \equiv \sigma_{sym} = \sigma_{xx} = \sigma_{yy}$.

Now we can substitute the expression for the isotropic (symmetric) part of the effective 
conductivity found up to the sixth order of the theory of perturbation
\begin{equation}
\sigma_{eff} = 1 - \frac{\delta^2}{3} - \frac{\delta^4}{12} - \frac{17\delta^6}{432} - \cdots 
\label{expansion}
\end{equation}
into the three-color Bruggeman equation (\ref{three-col}) and to see that it is satisfied. It gives us some grounds to think that this equation describes exactly    
the effective conductivity for the chaotic three-color distributions
of conductivity on the plane just like it does for the two-color distributions,
where the formula of Bruggeman (\ref{two-col}) coincides with the exact formula of Dykhne.

It is necessary to stress that for the chessboard with 
conductivity described by function $\alpha(x,y)$ from Eq. 
(\ref{three-color}) there is an anisotropic term already in 
the second order of perturbation theory, which corresponds to the following diagram 

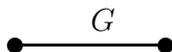
\begin{picture}(40,20)
\put(10,10){\circle*{2}}
\put(10,10){\line(4,0){20}}
\put(20,12){$G$}
\put(30,10){\circle*{2}}
\end{picture}

and is equal to
\begin{equation}
\sigma_{xy} =  
-\frac{486\delta^2}{(2\pi)^4}
\sum_{k,l=-\infty}^{\infty}\frac{1}
{(3k+2)(3l+1)((3k+2)^2+(3l+1)^2)}\approx 0.06\delta^2.
\label{anys3}
\end{equation}

We would like to stress that for isotropic structures the conductivity in the second order is purely local and has universal form which does not depend on its
particular structure. In anisotropic structures already in the second order we encounter non-local propagators and the results is not universal.

Now let us consider another three-color chessaboard with nonequal weights, which however is an isotropic one in contrast to the preceding one. 
The conductivity here is given by the formulae

\begin{eqnarray}
&&\alpha(x,y) =  0,\  if \   0<x<\pi,\ 0<y<\pi ; \nonumber \\    
&&\alpha(x,y) = + \delta,\  if \   \pi<x<2\pi,\ 0<y<\pi ; \nonumber \\    
&&\alpha(x,y) = - \delta,\  if \   0<x<\pi,\ \pi<y<2\pi ; \nonumber \\    
&&\alpha(x,y) =  0,\   if \   \pi<x<2\pi,\ \pi<y<2\pi.
\label{alpha3}
\end{eqnarray}

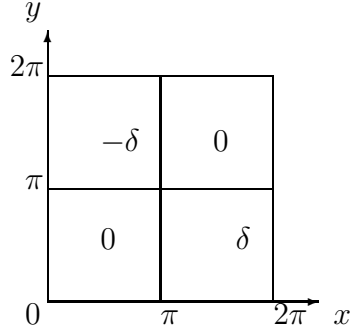
\begin{picture}(43,43)
\put(5,5){\vector(2,0){36}}
\put(5,5){\vector(0,2){36}}
\put(2,2){$0$}
\put(20,2){$\pi$}
\put(35,2){$2\pi$}
\put(5,20){\line(2,0){30}}
\put(5,35){\line(2,0){30}}
\put(20,5){\line(0,2){30}}
\put(35,5){\line(0,2){30}}
\put(2,20){$\pi$}
\put(0,35){$2\pi$}
\put(43,2){$x$}
\put(2,43){$y$}
\put(12,12){$0$}
\put(30,12){$\delta$}
\put(12,25){$-\delta$}
\put(27,25){$0$}
\end{picture}

Though this chessboard is a three-color one, it possesses all the symmetries of the two-color chessboard and is an isotropic one.

Non-zero Fourier coefficients have the following form:
\begin{equation}
\alpha_{2k+1,0} = \frac{1}{i\pi(2k+1)},
\label{four1}
\end{equation}
\begin{equation}
\alpha_{0,2l+1} = \frac{1}{i\pi(2l+1)}.
\label{four2}
\end{equation}
It is convenient to denote these Fourier coefficients by different vertices:
the coefficient from the formula (\ref{four1}) by a black circle and that one
from formula (\ref{four2}) by a white circle. 

The contribution of the second order of the perturbation theory into effective conductivity is given by the diagram 

\begin{picture}(40,20)
\put(10,10){\circle*{2}}
\put(10,10){\line(4,0){20}}
\put(30,10){\circle*{2}}
\end{picture}

and is equal to
\begin{equation}
\sigma_{eff 2} = - \frac{\delta^2}{4},
\label{weigt}
\end{equation}

The diagrams of the fourth order have the following form

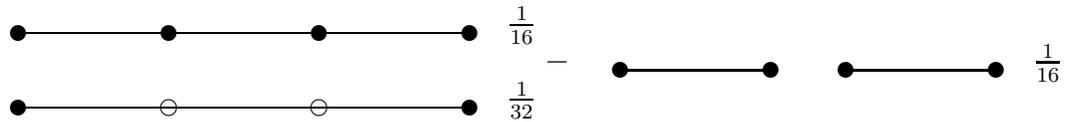
\begin{picture}(150,30)
\put(10,25){\circle*{2}}
\put(10,25){\line(4,0){20}}
\put(30,25){\circle*{2}}
\put(30,25){\line(4,0){20}}
\put(50,25){\circle*{2}}
\put(50,25){\line(4,0){20}}
\put(70,25){\circle*{2}}
\put(75,25){$\frac{1}{16}$}
\put(10,15){\circle*{2}}
\put(10,15){\line(4,0){20}}
\put(30,15){\circle{2}}
\put(30,15){\line(4,0){20}}
\put(50,15){\circle{2}}
\put(50,15){\line(4,0){20}}
\put(70,15){\circle*{2}}
\put(75,15){$\frac{1}{32}$}
\put(80,20){$-$}
\put(90,20){\circle*{2}}
\put(90,20){\line(4,0){20}}
\put(110,20){\circle*{2}}
\put(120,20){\circle*{2}}
\put(120,20){\line(4,0){20}}
\put(140,20){\circle*{2}}
\put(145,20){$\frac{1}{16}$}
\end{picture}

Thus, the fourth-order contribution is
\begin{equation}
\sigma_{eff4} = -\frac{\delta^4}{32}.
\label{weight1}
\end{equation}

The Bruggeman equation for this structure could be written as \cite{Brug}
\begin{equation}
2 \frac{\sigma_{eff} - \sigma_1}{\sigma_{eff} + \sigma_1} +
\frac{\sigma_{eff} - \sigma_2}{\sigma_{eff} + \sigma_2} +
\frac{\sigma_{eff} - \sigma_3}{\sigma_{eff} + \sigma_3} = 0,
\label{Brug2}
\end{equation}
or equivalently
\begin{equation}
\sigma_{eff}^3 + \frac{\sigma_2+\sigma_3}{2}\sigma_{eff}^2 - 
\frac{\sigma_1(\sigma_2+\sigma_3)}{2}\sigma_{eff} - \sigma_1\sigma_2\sigma_3 = 0.
\label{Brug3}
\end{equation}
For our case when,
$$
\sigma_1 = 1, \ \sigma_2 = 1 + \delta, \sigma_3 = 1 - \delta,
$$
one can easily get
\begin{equation}
\sigma_{eff} = 1 - \frac{\delta^2}{4} - \frac{\delta^4}{16} - \cdots.
\label{Brug4}
\end{equation}
It is obvious that in the fourth order of the perturbation theory there is a
contradiction between the results of perturbative calculations (\ref{weight1}) and those following from the Bruggeman equation (\ref{Brug4}). 

\section*{ACKNOWLEDGEMENTS}
This work was partially supported by RFBR via grant No 99-02-18409. 
Work of A.K. was also supported by CARIPLO 
Scientific Foundation. 
We are grateful to A.M. Dykhne, D.J. Bergman and L.G. Fel who have 
introduced  us to the problem of effective conductivity of composite 
systems. 
We would thank J.B. Keller for the interest in our work.
We are grateful also to A.S. Ioselevich, 
I.V. Kolokolov, V.G. Marikhin, A.V. Marshakov, A.Yu. Morozov, 
A.A. Rosly and  A.V. Zabrodin for the useful remarks.

\end{document}